\documentclass[prd,preprintnumbers,nofootinbib,aps]{revtex4}
\usepackage{epsfig,amsmath,amssymb}
\usepackage{mathrsfs}
\usepackage[usenames,dvipsnames]{color}
\usepackage[pagebackref=false, colorlinks=false]{hyperref}

\begin{document}

\title{Probing the post-Minkowskian approximation using recursive addition of self-interactions}

\author{Soumendra Kishore Roy}
\email{soumendrakishoreroy09@gmail.com} 
\affiliation{Department of Physics, Presidency University, Kolkata 700073, 
India.}

\author{Ratna Koley}
\email{ratna.physics@presiuniv.ac.in} 
\affiliation{Department of Physics, Presidency University, Kolkata 700073, 
India.}  

\author{Parthasarathi Majumdar}
\email{bhpartha@gmail.com}
\affiliation{School of Physical Sciences, Indian Association for the Cultivation of Science, Kolkata 700032, India }

\begin{abstract}
    We address the problem of deriving the post-Minkowskian approximation, widely used in current gravitational wave literature by investigating a possible deduction out of  the recursive N\"other coupling approach, from the Pauli-Fierz spin-2 theory in flat spacetime. We find that this approach yields the post-Minkowskian approximation correctly to the first three orders, without invoking any weak-field limit of general relativity. This connection thus establishes that the post-Minkowskian approximation has a connotation independent of a weak-field expansion of general relativity, which is the manner usually presented in the literature. As a consequence, a link manifests between the  recursive N\"other coupling  approach to  deriving general relativity from a linear spin-2 theory in flat spacetime and theoretical analyses of recent detection of gravitational wave events.
\end{abstract}

\maketitle

\section{Introduction}

For the computation of asymptotic waveforms for specific types of gravitating sources, various approximation methods are employed for the solution of Einstein's equation. The post-Minkowskian approximation combined with the post-Newtonian approximation is such a method applicable where the gravitational field can be {\it assumed} to be weak. The post-Minkowskian approximation is generally used for asymptotically flat spacetimes, far from the gravitating source. The Einstein-Hilbert action is approximated in terms of the perturbation series around the flat Minkowski metric by taking the contravariant metric {\it density}, and expanding the Einstein-Hilbert action (or equivalently, the Einstein equation) in powers of its departure from the Minkowski metric (in powers of the gravitational constant $G$). Each order of the perturbation of the post-Minkowskian series satisfies the inhomogeneous wave equation whose solution is given by some sort of multipole expansion, if the source is slowly moving. To ascertain the form of the gravitational field distribution in spacetime for a specific kind of source, the post-Minkowskian approximation is combined with the post-Newtonian approximation, in a common zone where both approximations are valid \cite{Blanchet}- \cite{Blanchet1}. The post-Minkowskian approximation is valid for all velocities under the weak field approximation of general relativity. However, post-Newtonian series takes gravity to be Newtonian in the zeroth order of the series and it is essentially a $v/c$ expansion for {\it every power of $G$}. Due to its non-relativistic structure, such approximation scheme works fine until the velocity of source becomes comparable to light \cite{Blanchet}. This methodology has evolved tremendously  since the early days \cite{Clifford}, and recently has been shown to be very useful by yielding many observational results \cite{Verify}-\cite{Verify8}.

Despite the proximity of the post-Minkowskian approximation to the weak field expansion of general relativity, there is as of now no systematic `bottoms-up' approach of {\it deriving} the leading nonlinear interactions, starting from only a special relativistic theory, albeit a free field theory. The fate of the foundational local invariance principles of general relativity, namely general coordinate invariance and local Lorentz invariance, is somewhat uncertain in the post-Minkowskian expansion : neither invariance principle is rigidly retained or crucially used in that expansion. In this sense, the post-Minkowskian expansion is somewhat ad hoc foundationally, notwithstanding its methodological utility in gravitational wave signal processing and source modelling. The issue we address in this paper is : can we formulate the post-Minkowskian approximation, starting not with the full nonlinear general relativity theory, but with a special relativistic theory, and systematically deriving nonlinear interactions which agree, order by order, with the post-Minkowskian approximation expansion of general relativity, without having to invoke any physical restriction to {\it weak} fields ?   

There is an additional physical motivation behind this work: there is a large body of work on perturbative graviton scattering where it  been shown \cite{QG}-\cite{QG7} how the classical limit of the graviton scattering amplitude reduces to the post-Minkowskian result. The question is : is (perturbative) quantization of linearized gravity germane to this derivation, or is there a purely classical approach which yields the same results ?   

The derivation of full nonlinear general relativity theory, starting from a special relativistic field theory, is a century-old question in the field of theoretical physics. Einstein's own construction of general relativity is based on the {\it minimal coupling} prescription coming from the equivalence principle, where a partial coordinate derivative is supplanted by a new derivative {\it covariant} under general coordinate transformations. This inspired Yang and Mills to formulate a locally gauge-invariant field theory under a  non-Abelian group, from the action with global gauge invariance \cite{Yang} under the same group. The formal derivation of the principle of gauge or general covariance from the Lorentz invariant Fierz-Pauli action \cite{Fierz} has been attempted by several authors \cite{Author}-\cite{Author2}, but with varying degrees of success. One of the more successful derivations of the minimal coupling prescription, both in the case of Yang-Mills theory and, to some extent, for general relativity, has been done by Deser \cite{Deser}. The approach consists in identifying the interaction between the field and the N\"other current (as a result of global translational symmetry in deriving GR, and global non-Abelian symmetry for Yang-Mills theory) as the source of the next order field equation.

The recursive N\"other coupling as the source term in the free field action successfully generates the Yang-Mills theory (both with and without matter) in finite steps \cite{Yang1}-\cite{Yang5}. This essentially implies that non-Abelian {\it local gauge} invariance is basically a {\it derived} concept, and all field interactions of the standard strong-electroweak theory can be derived from an Abelian gauge theory with a global non-Abelian invariance, by recursive N\"other coupling \cite{bmm18}. However, general relativity needs an infinite number of such non-linear interactions (in terms of the field variable) to give the full generally covariant theory. It has been argued  \cite{Paddy} there is an issue of convergence : does the infinite series {\it converge} to general relativity ? It has been  shown \cite{Deser} that a first-order formulation partially mitigates the situation, while performing the recursion. This can generate the bulk Einstein-Hilbert term \cite{Paddy} in the action, modulo spacetime derivatives which contribute at the boundary; however, for the Einstein equation, this is deemed sufficient \cite{Deser1}.

In this paper, we establish the connection between the recursive N\"other coupling and the post-Minkowskian expansion up to the third-order perturbation. We explicitly show how consistency of the recursive N\"other coupling demands the non-linear interaction terms to generate order-by-order post-Minkowskian results up to a specific order, starting from the Fierz-Pauli action. Our paper makes this connection {\it without} having to resort to quantizing the theory, contrary to ref. \cite{QG}-\cite{QG7}. To reiterate, our approach does not rely on the strong field aspects of general relativity. This is novel because it generates the far-field expansion of the Einstein field equation starting from the action of the gravitational wave which is globally Poincar\'e-invariant instead of having general coordinate invariance. This correspondence also enables us to indirectly establish the practical utility of the recursive N\"other coupling in analyses of binary merger events leading to observable gravitational wave signals. However, we hasten to add that our approach does not depend on any fewer fundamental assumptions than either the standard Post-Minkowskian paradigm, or any possible derivation from eikonal graviton scattering in the appropriate limit. Our work is therefore an alternative, but {\it equivalent} approach to those assays.

Section II briefly discusses the conventional way of doing post-Minkowskian approximation from general relativity. In section III, we extract the physical part of the field variable in the Fierz-Pauli action with the help of a {\it projection operator}, derived in \cite{ar19}. This is an alternative to the conventional approach of gauge-fixing and  has been shown to be useful in classical electrodynamics \cite{mr2019}-\cite{rma18}. In section IV we derive the post-Minkowskian series of GR by recursively adding the self-interaction term (function of the physical part of the field) from the linearized theory of gravity based on a purely classical viewpoint. In section V we conclude by discussing this correspondence in modified gravity theories and the future prospect of this work in numerical relativity.

\section{Review of Post-Minkowskian Approximation}

Einstein's equation for a spacetime with energy-momentum tensor $T^{ab}$ and metric $g_{ab}$ is given by,
\begin{equation}\label{Einstein}
    G^{ab} = 8\pi GT^{ab}
\end{equation}
Here, velocity of light $c=1$. $G^{ab}=R^{ab}-\frac{1}{2}g^{ab}R$ is the Einstein tensor which is function of metric $g_{ab}$, and its first and second derivative. Now, let's define a new field variable $h^{ab}$ as,
\begin{center}
    $h^{ab}=\sqrt{-g}g^{ab}-\eta ^{ab}$
\end{center}
where $g=det(g_{ab})$ and $\eta^{ab}$ is the Minkowskian metric. The definition of $h^{ab}$ is not contrary to general relativity as we have not put any constraint on $h^{ab}$. The field $h^{ab}$ is solely defined by the geometry itself \cite{Blanchet1}. The general coordinate invariance of the spacetime allows us to choose a coordinate system. Here, the de-Donder frame defined by,
\begin{equation}\label{Gauge}
      \partial_a h^{ab}=0
\end{equation}
is chosen. The advantage of this coordinate system lies in the match with transverse gauge condition of the linear field in flat spacetime. In this gauge, Einstein equation \eqref{Einstein} for $h^{ab}$ becomes,
\begin{equation}\label{Pm}
    \Box h^{ab}=\tau ^{ab}(T^{ab}, h^{ab}, \partial_c h^{ab}, \partial_c \partial_d h^{ab})
\end{equation}
where $\Box=\partial_a \partial^a$ defined on the Minkowskian metric $\eta^{ab}$ and $\tau^{ab}(T^{ab}, h^{ab}, \partial_c h^{ab}, \partial_c \partial_d h^{ab})$ is given by,
\begin{equation}\label{tau}
    \tau^{ab}(T^{ab}, h^{ab}, \partial_c h^{ab}, \partial_c \partial_d h^{ab})=|g|T^{ab}+\frac{1}{16\pi G}\Lambda^{ab}(h^{ab}, \partial_c h^{ab}, \partial_c \partial_d h^{ab})
\end{equation}
The form of pseudo tensor $\Lambda^{ab}(h^{ab}, \partial_c h^{ab}, \partial_c \partial_d h^{ab})$ is as follows. 
\begin{equation}\label{Pseudo}
    \Lambda^{ab}(h^{ab}, \partial_c h^{ab}, \partial_c \partial_d h^{ab})=-h^{cd}\partial_c \partial_d h^{ab}+\partial_c h^{ad} \partial_d h^{bc}+\frac{1}{2}g^{ab}g_{cd}\partial_e h^{cf} \partial_f h^{de}-g^{ac}g_{de}\partial_f h^{be}\partial_c h^{df}-g^{bc}g_{de}\partial_f h^{ae}\partial_c h^{df}
\end{equation}
\begin{center}
    $+g_{cd}g^{ef}\partial_e h^{ac}\partial_f h^{bd}+\frac{1}{8}(2g^{ac}g^{bd}-g^{ab}g^{cd})(2g_{ef}g_{pq}-g_{ep}g_{fq})\partial_c h^{eq}\partial_d h^{fp}$
\end{center}
Note that, $\Lambda^{ab}$ is not only the function of $h^{ab}$ and its derivatives but also function of the metric. 
According to the definition of $h^{ab}$, $\Lambda^{ab}$ contains all powers of $h^{ab}$, starting with bilinear terms. Hence in the weak field limit, the expression of $\Lambda^{ab}$ is written as,
\begin{center}
    $\Lambda^{ab}=N^{ab}(h,h)+M^{ab}(h,h,h)+O(h^4)$
\end{center}
The expressions of $N^{ab}(h,h)$ and $M^{ab}(h,h,h)$ can be found in \cite{Blanchet}, \cite{Blanchet1}. In this paper, we will later derive their expressions from recursive N\"other current. Now, in a more formal language, the field variable $h^{ab}$ is a series sum of perturbations in the weak field limit,
\begin{center}
    $h^{ab}=Gh^{ab}_{(1)}+G^2h^{ab}_{(2)}+G^3h^{ab}_{(3)}+O(h^4)$
\end{center}
where $h^{ab}_{(n)}$ is found by recursively solving the field equations,
\begin{equation}
    \Box h^{ab}_{(1)}=0
\end{equation}
\begin{equation}
    \Box h^{ab}_{(2)}=N^{ab}(h_{(1)},h_{(1)})
\end{equation}
\begin{equation}
    \Box h^{ab}_{(3)}=N^{ab}(h_{(2)},h_{(1)})+N^{ab}(h_{(1)},h_{(2)})+M^{ab}(h_{(1)},h_{(1)},h_{(1)})
\end{equation}
The above field equations of $h^{ab}_{(n)}$ are known as the post-Minkowskian field equations and the series of $h^{ab}$ in terms of $h^{ab}_{(n)}$ is the post-Minkowskian series. Physically, $h^{ab}_{(n)}$ is the $n^{th}$ order perturbation propagating through the flat spacetime. The derivation presented here comes from approximating general relativity in the weak field limit, and hence it relies on the general coordinate invariance. However, in the view of flat spacetime Pauli-Fierz spin $2$ theory, post-Minkowskian perturbations should be understood in terms of Lorentz invariance without invoking general coordinate invariance. 
This is the main goal of the next two sections.  

\section{The Physical Part of the Field in the Fierz-Pauli Action}

The Fierz-Pauli action for the field variable $h_{ab}({\bf x})$, characterizing a massless, spin 2 field, is given in the coordinates ${\bf x}$ (boldface letter denotes the four vector in the compact form and the letter with Latin indices denotes the components), 
\begin{equation}\label{Action}
S = \frac{1}{64 \pi G} \int _{\upsilon} d^4x (-\partial_a h_{bc}\partial^a h^{bc}+\partial_a h^b_{b}\partial^a h^c_{c}-2 \partial_a h^a_{c} \partial^c h^b_{b}+2\partial_a h^a_{c} \partial_b h^{bc})
\end{equation}
where $\upsilon$ is the four volume under consideration, $G$ is the Newton's gravitational constant and the velocity of light, $c=1$. $S$ uniquely describes the action of a symmetric second rank tensor field within the domain of Lorentz invariance and the action can be constructed without any prior knowledge of the principle of general covariance \cite{Fierz}. The corresponding equation of motion of $h^{ab}({\bf x})$ in the presence of the self-interacting source $T^{ab}(h^{pq},\partial^r h^{pq})$ is-
\begin{equation}\label{heqn}
    \Box h^{ab} - \eta^{ab} \Box h + \partial^a \partial^b h + \eta^{ab} \partial_c \partial_d h^{cd} - 2\partial^{(a}\partial_c h^{b)c}=T^{ab}(h^{pq},\partial^r h^{pq})
\end{equation}
where $h = \eta_{ab} h^{ab}$ and the symmetric part of second rank Lorentz tensor is given by $A^{(ab)}=\frac{1}{2}(A^{ab}+A^{ba})$. 

The transformation $\Bar{h}^{ab}=h^{ab}-\frac{1}{2}\eta^{ab}h$ changes the equation \eqref{heqn} to,
\begin{equation}\label{hbareqn}
    \Box \Bar{h}^{ab} - \partial^a \partial_c \Bar{h}^{bc} - \partial^b \partial_c \Bar{h}^{ac} + \eta^{ab} \partial_c \partial_d \Bar{h}^{cd} = \Bar{T}^{ab}(\Bar{h}^{pq},\partial^r \Bar{h}^{pq})
\end{equation}
Fourier transform of the equation \eqref{hbareqn} gives,
\begin{center}
    $\textbf{k}^2 \Tilde{\Bar{h}}^{ab} - k^a k_c \Tilde{\Bar{h}}^{bc} - k^b k_c \Tilde{\Bar{h}}^{ca} + \eta^{ab}k_c k_d \Tilde{\Bar{h}}^{cd} = \Tilde{\Bar{T}}^{ab}$
\end{center}
where $\Tilde{\Bar{h}}^{ab}$ and $\Tilde{\Bar{T}}^{ab}$ are the four-Fourier transform of $\Bar{h}^{ab}$ and $\Bar{T}^{ab}$. In the presence of matter ($\Tilde{\Bar{T}}^{ab} \neq 0$), $\textbf{k}^2$ is not $0$ and this simplifies the equation of motion to,
\begin{equation}\label{hPeqn}
    \textbf{k}^2 P^{ab}_{cd} \Tilde{\Bar{h}}^{cd} = \Tilde{\Bar{T}}^{cd} 
\end{equation}
with,
\begin{equation}\label{Projection}
    P^{ab}_{cd} = \delta^a_{(c} \delta^b_{d)} - \frac{k^a k_{(d}}{\textbf{k}^2}\delta^b_{c)} - \frac{k^b k_{(c}}{\textbf{k}^2}\delta^a_{d)} + \eta^{ab}\frac{k_c k_d}{\textbf{k}^2} 
\end{equation}
Now $P^{ab}_{cd}$ satisfies the property of the projection operator,
\begin{center}
$P^{ab}_{cd} P^{cd}_{ef} = P^{ab}_{ef}$
\end{center}
which shows one of the eigen value of $P^{ab}_{cd}$ is zero and henceforth projection operator is non-invertible. Here the conventional approach is choosing a gauge and solve for the field $\Bar{h}^{cd}$. This imposes a choice on $\Bar{h}^{cd}$ and the solution completely depends on that choice. In this paper we take rather an unconventional way of dealing this non-invertibility- we identify $P^{ab}_{cd} \Bar{h}^{cd}$ as the physical part of $\Bar{h}^{cd}$, only which governs the equation of motion \eqref{hPeqn}. The rest of $\Bar{h}^{cd}$ is redundant for the dynamics. This method is described in detail at \cite{ar19}. Note that, in case of electrodynamics, the problem is similar and the same technique is also applicable there \cite{mr2019}, \cite{rma18}. Hence the physical degrees of freedom of the field $\Bar{h}^{cd}$ is obtained by the projected field,
\begin{equation}\label{Projected1}
    \Bar{h}^{ab}_{(P)}=P^{ab}_{cd} \Bar{h}^{cd}
\end{equation}
The gauge invariance of $\Bar{h}^{cd}_{(P)}$ is automatically guaranteed w.r.t a new gauge variable. The transversality of the projected field $\Bar{h}^{cd}_{(P)}$ is also satisfied,
\begin{center}
    $k_b \Tilde{\Bar{h}}^{ab}_{(P)}=k_b P^{ab}_{cd} \Tilde{\Bar{h}}^{cd}=0$ as $k_b P^{ab}_{cd}=0$.
\end{center}
i.e.,
\begin{equation}\label{Transverse}
    \partial_b \Bar{h}^{ab}_{(P)}=0
\end{equation}
The Fierz-Pauli action is written in terms of the projected field $h^{ab}_{(P)}$ as,
\begin{equation}\label{Action1}
    S = \frac{1}{64 \pi G} \int _{\upsilon} d^4x (-\partial^a \Bar{h}^{bc}_{(P)} \partial_a h_{bc(P)} + \frac{1}{2} \partial_c \Bar{h}_{(P)} \partial^c \Bar{h}_{(P)})
\end{equation}
In the context of the gravitational wave (which is the physical interpretation of $h^{ab}(\textbf{x})$), the transverse-traceless projection is frequently used \cite{Blanchet1}, \cite{Primordialgw}. But the purpose of using the projection operator in those literature is just the mathematical convenience, the physicality of the projected field is not given any stress. 
The novelty of our approach is that, we generate the entire post-Minkowskian series only from the projected field and hence there is no question of the gauge ambiguity in our formalism.

\section{post-Minkowskian Expansion in Terms of the Recursive N\"other Coupling}
The Fierz-Pauli action \eqref{Action1} is equivalent to a new action $S_0$ on addition of a boundary term $S^{\prime}$, where $S^{\prime}$ and $S_0$ are given by,

\begin{equation}\label{ActionBoundary}
    S^{\prime} = \frac{1}{32 \pi G} \int _{\upsilon} d^4x [\partial_a (\partial_e \Bar{h}^{ab}_{(P)} \Bar{h}^{e}_{b(P)}) - \partial_e (\partial_a \Bar{h}^{ab}_{(P)} \Bar{h}^{e}_{b(P)})]
\end{equation}
and,
\begin{equation}\label{Action2}
    S_0 = \frac{1}{32 \pi G} \int _{\upsilon} d^4x M^{ef}_{abcd} (\eta^{mn}) \partial_e \Bar{h}^{ab}_{(P)} \partial_f \Bar{h}^{cd}_{(P)}
\end{equation}
where,
\begin{equation}\label{M}
    M^{ef}_{abcd} (\eta^{mn}) = -\frac{1}{2}(\eta_{ac} \eta_{bd} - \frac{1}{2} \eta_{ab} \eta_{cd}) \eta^{ef} + \eta_{bc} \delta^e_d \delta^f_a
\end{equation}
The boundary action \eqref{ActionBoundary} has zero contribution under the no field exchange condition at the boundary of $\upsilon$.
Now, let us find the Belinfante energy-momentum tensor of the action $S_0$. 
The energy-momentum tensor $B_{pq}^{(1)}$ of the action \eqref{Action1} is obtained by writing it in a spacetime 
with auxiliary metric $\gamma^{ab}$ and then taking the $\gamma ^{ab} \rightarrow \eta ^{ab}$ limit,
\begin{center}
    $B_{pq}^{(1)}=\frac{2}{\sqrt{-\gamma}}\frac{\delta S_0 [\gamma^{ab}, \partial^c \gamma^{ab}]}{\delta \gamma^{pq}}|_{\gamma^{pq}=\eta^{pq}}$
\end{center}
\begin{center}
    $B_{pq}^{(1)}=\frac{2}{\sqrt{-\gamma}} \big[\frac{\partial {\cal L}}{\partial \gamma^{pq}}-\partial_r \big(\frac{\partial {\cal L}}{\partial(\partial_r\gamma^{pq})} \big) \big]|_{(\gamma^{pq}=\eta^{pq})}$
\end{center}
The determinant of the covariant form of the metric is denoted by 
$\gamma$ and $\cal L$ is the Lagrangian density of $S_0$. The second part of $B_{pq}^{(1)}$ can be dropped by the ambiguity of the energy-momentum tensor on the addition of a first-order derivative of a third rank two indices antisymmetric tensor. 
For the given action, $B^{(1)}_{pq}$ is explicitly computed as,
\begin{center}
    $B^{(1)}_{pq}=\frac{1}{32 \pi G} \big[- \frac{1}{2} \big( \eta_{p(c} \eta_{qa)} \eta_{bd} - \frac{1}{2} \eta_{p(c} \eta_{qd)} \eta_{ab} \big)\eta^{ef}-\frac{1}{2}\big( \eta_{ca} \eta_{bd} -\frac{1}{2} \eta_{ab} \eta_{cd} \big) \delta^{(e}_p \delta^{f)}_q +\eta_{p(b} \eta_{qc)} \delta^e_d \delta^f_a -\eta_{p(c} \eta_{qa)} \delta^e_b \delta^f_d \big] \partial_e \Bar{h}^{ab}_{(P)} \partial_f \Bar{h}^{cd}_{(P)}$.
\end{center}

The self-interaction requires $B^{(1)}_{pq}$ to be the source term in the next order field equation. Correspondingly, the action $S_0$ gets modified to $S_{1,tot}$ by the coupling of $B^{(1)}_{pq}$ with $\Bar{h}^{pq}_{(P)}$,
\begin{equation}\label{Action3}
    S_{1,tot}=S_0+S_1
\end{equation}
where,
\begin{center}
    $S_1=\int_{\upsilon} d^4x ~B^{(1)}_{pq} \Bar{h}^{pq}_{(P)}$
\end{center}
i.e.
\begin{equation}\label{Action4}
    S_1=\frac{1}{32 \pi G}\int _{\upsilon} d^4x \big[- \frac{1}{2} \big( \Bar{h}_{ca (P)} \eta_{bd} - \frac{1}{2}\Bar{h}_{cd (P)}\eta_{ab} \big)\eta^{ef}-\frac{1}{2}\big( \eta_{ca} \eta_{bd} -\frac{1}{2} \eta_{ab} \eta_{cd} \big) \Bar{h}^{ef}_{(P)}+\Bar{h}_{bc (P)}\delta^e_d \delta^f_a -\Bar{h}_{ca (P)} \delta^e_b \delta^f_d \big] \partial_e \Bar{h}^{ab}_{(P)} \partial_f \Bar{h}^{cd}_{(P)}
\end{equation}

On extremisation, $S_{1,tot}$ gives the field equation with quadratic self-interaction,
\begin{equation}\label{EOM2}
    \Box h^{ab}=N^{ab}(\Bar{h}^{pq}_{(P)},\Bar{h}^{rs}_{(P)})
\end{equation}
where the self interaction is as follows,
\begin{center}
    $N^{ab} (\Bar{h}^{pq}_{(P)},\Bar{h}^{rs}_{(P)})
    =-\Bar{h}^{cd}_{(P)} \partial_c \partial_d \Bar{h}^{ab}_{(P)} + \frac{1}{2} \partial^a \Bar{h}_{cd (P)} 
    \partial^b \Bar{h}^{cd}_{(P)}-\frac{1}{4}\partial^a \Bar{h}_{(P)} \partial^b \Bar{h}_{(P)}-2 \partial^{(a}
    \Bar{h}_{cd (P)} \partial^c \Bar{h}^{bd)}_{(P)}$
\end{center}
\begin{equation}\label{Nab}
     + \partial_d \Bar{h}^{ca}_{(P)} (\partial^d \Bar{h}^b_{c (P)}+ \partial_c \Bar{h}^{bd}_{(P)})+ 
     \eta^{ab} \left[\frac{1}{4} \partial_c \Bar{h}_{de (P)} \partial^c \Bar{h}^{de}_{(P)}+
     \frac{1}{8} \partial_c \Bar{h}_{(P)} \partial^c \Bar{h}_{(P)}+ \frac{1}{2} \partial_c \Bar{h}_{de (P)} 
     \partial^d \Bar{h}^{ce}_{(P)} \right]
\end{equation}
For the next order field equation is with the cubic self-interaction, the energy-momentum tensor of $S_1$ couples with the field variable 
$\Bar{h}^{ab}_{(P)}$. The corresponding total action $S_{2,tot}$ is given in terms of the correction of $S_{1,tot}$ by the action $S_2$. Where $S_2$ 
is given as follows
%\begin{center}
\begin{eqnarray}\label{Action5}
    %$
    S_2 &=&\frac{1}{32 \pi G}\int_{\upsilon} d^4x [-(\Bar{h}_{ar (P)} \Bar{h}^r_{c (P)} \eta_{bd}+ \Bar{h}_{dr (P)} \Bar{h}^r_{b (P)} \eta_{ca}+ \Bar{h}_{ca (P)} \Bar{h}_{bd (P)}-
    %$
%\end{center}
%\begin{center}
    %$
    \frac{1}{2}(\Bar{h}_{br (P)} \Bar{h}^r_{a (P)} \eta_{cd} 
    + \Bar{h}_{dr (P)} \Bar{h}^r_{c (P)} \eta_{ab} \\ \nonumber 
    &+& \Bar{h}_{ab (P)} \Bar{h}_{cd (P)}))\partial_e \Bar{h}^{ab (P)} \partial_f \Bar{h}^{cd}_{(P)} \eta^{ef}- (\Bar{h}_{ca (P)} \eta_{bd}+ \eta_{ca} \Bar{h}_{bd (P)}-
    %$
%\end{center}
%\vspace{-0.5cm}
%\begin{equation}\label{Action5}
     \frac{1}{2}(\Bar{h}_{ab (P)} \eta_{cd}+ \eta_{ab} \Bar{h}_{cd (P)}))\partial_e \Bar{h}^{ab}_{(P)} \partial_f \Bar{h}^{cd}_{(P)} \Bar{h}^{ef}_{(P)}
      \\ \nonumber &+& 2 \Bar{h}^r_{c (P)} \Bar{h}_{pb (P)} \partial_d \Bar{h}^{ab}_{(P)} 
    \partial_a \Bar{h}^{cd}_{(P)}  - 2 \Bar{h}^p_{a (P)} \Bar{h}_{pc (P)} \partial_b \Bar{h}^{ab}_{(P)} \partial_d \Bar{h}^{cd}_{(P)}]
%\end{equation}
\end{eqnarray}
Now the total action 
\begin{equation}\label{Action6}
    S_{2,tot}=S_0+S_1+S_2
\end{equation}
on extremisation gives the field equation with the quadratic and cubic self-interactions as the source terms
\begin{equation}\label{EOM3}
    \Box \Bar{h}^{ab}_{(P)}=N^{ab} (\Bar{h}^{pq}_{(P)}, \Bar{h}^{rs}_{(P)})+ M^{ab}(\Bar{h}^{pq}_{(P)},\Bar{h}^{rs}_{(P)}, \Bar{h}^{uv}_{(P)}).
\end{equation}
We found $N^{ab} (\Bar{h}^{pq}_{(P)}, \Bar{h}^{rs}_{(P)})$ in \eqref{Nab} and $M^{ab}$ is given by
\begin{center}
    $M^{ab}(\Bar{h}^{pq}_{(P)},\Bar{h}^{rs}_{(P)}, \Bar{h}^{uv}_{(P)}) = -\Bar{h}^{cd}_{(P)} (\partial^a \Bar{h}_{ce (P)} \partial^b \Bar{h}^e_{d (P)}+ \partial^e \Bar{h}^a_{c (P)} \partial^e \Bar{h}^b_{d (P)}- \partial_c \Bar{h}^a_{e (P)} \partial_d \Bar{h}^{be}_{(P)}) +$
\end{center} 
\begin{center} 
    $\Bar{h}^{ab}_{(P)} [-\frac{1}{4}\partial_c \Bar{h}_{de (P)} \partial^c \Bar{h}^{de}_{(P)} +\frac{1}{8}\partial_c \Bar{h}_{(P)} \partial^c \Bar{h}_{(P)}+ \frac{1}{2}\partial_c \Bar{h}_{de (P)} \partial^d \Bar{h}^{ce}_{(P)}]+$
\end{center}
\begin{center}
    $\frac{1}{2} \Bar{h}^{cd}_{(P)} \partial^{(a} \Bar{h}_{cd (P)} \partial^{b)} \Bar{h}_{(P)} + 2 \Bar{h}^{cd}_{(P)} \partial_e \Bar{h}^{(a}_{c (P)} \partial^{b)} \Bar{h}^e_{d (P)}+ \Bar{h}^{c(a}_{(P)} ( \partial^{b)} \Bar{h}_{de (P)} \partial_c \Bar{h}^{de}_{(P)}- 2 \partial_d \Bar{h}^{b)}_{e (P)} \partial_c \Bar{h}^{de}_{(P)}- \frac{1}{2} \partial^{b)} \Bar{h}_{(P)} \partial_c \Bar{h}_{(P)})+$
\end{center}
\vspace{-0.5cm}
\begin{equation}\label{Mab}
      \eta^{ab}[\frac{1}{8} \Bar{h}^{cd}_{(P)} \partial_c \Bar{h}_{(P)} \partial_d \Bar{h}_{(P)}- \frac{1}{4} \Bar{h}^{cd}_{(P)} \partial_e \Bar{h}_{cd (P)} \partial^e \Bar{h}_{(P)}- \frac{1}{2} \Bar{h}^{cd}_{(P)} \partial_e \Bar{h}_{cf (P)} \partial^f \Bar{h}^e_{d (P)}+ \frac{1}{2} \Bar{h}^{cd}_{(P)} \partial_e \Bar{h}^f_{c (P)} \partial^e \Bar{h}_{df (P)}] 
\end{equation}
Both $N^{ab}$ and $M^{ab}$ satisfy the transversality condition by the equations of motions \eqref{EOM2} \& \eqref{EOM3},
\begin{center}
    $\partial_a N^{ab} = 0 ~~~ \partial_a M^{ab} = 0$
\end{center}
The same process continues to include any order self-interaction in the field equation. To maintain a specific order, the field variable $\Bar{h}^{ab}_{(P)}$ is written in terms of the sums of the powers of Newton's gravitational constant $G$,
\begin{equation}\label{EOM5}
    \Bar{h}^{ab}_{(P)} = {\Sigma}^{\infty}_{n=1} G^n \Bar{h}^{ab}_{ (P) (n)}
\end{equation}
where $\Bar{h}^{ab}_{ (P) (n)}$ is the solution of a particular order self-interaction field equation. 
The equations of $\Bar{h}^{ab}_{(P) (n)}$ for $n = 1, 2, 3$ are given up to the cubic interaction as the source are respectively 
\begin{equation}\label{EOM6}
    \Box \Bar{h}^{ab}_{(P) (1)}=0,
\end{equation}
\begin{equation}\label{EOM7}
    \Box \Bar{h}^{ab}_{(P) (2)}=N^{ab}(\Bar{h}^{pq}_{(P) (1)},\Bar{h}^{rs}_{(P) (1)}),
\end{equation}
and,
\begin{equation}\label{EOM8}
    \Box \Bar{h}^{ab}_{(P) (3)}=N^{ab}(\Bar{h}^{pq}_{(P) (2)},\Bar{h}^{rs}_{(P) (1)})+  N^{ab}(\Bar{h}^{pq}_{(P) (1)},\Bar{h}^{rs}_{(P) (2)})+ M^{ab}(\Bar{h}^{pq}_{(P) (1)},\Bar{h}^{rs}_{(P) (1)}, \Bar{h}^{uv}_{(P) (1)}).
\end{equation}
The interaction terms $N^{ab}(\Bar{h}^{pq}_{(P)},\Bar{h}^{rs}_{(P)})$, and $M^{ab}(\Bar{h}^{pq}_{(P)},\Bar{h}^{rs}_{(P)}, \Bar{h}^{uv}_{(P)})$ are given in equations 
\eqref{Nab}, and \eqref{Mab}.

The de-Donder gauge has been chosen in deriving the post-Minkowskian approximation in \cite{Blanchet}, \cite{Blanchet2}, \cite{Blanchet1}. 
But the definition of the gauge fixed field in those literature matches with our definition of the projected field due to its transverse nature. 
Now the equations \eqref{EOM6}-\eqref{EOM8} are derived from the action of linearized gravity by recursively adding the self-interactions. 
These equations \eqref{EOM6}-\eqref{EOM8} are exactly same as found in the post-Minkowskian technique which is the approximation of the 
Einstein-Hilbert action in the $1^{st}$ and $2^{nd}$ post-Minkowskian order of $\Bar{h}^{ab}_{(P)}$ \cite{Blanchet}, \cite{Blanchet2}, 
\cite{Blanchet1}. The same job is done in \cite{Eikonal} by adding self-interacting gravitons in corresponding Feynman diagrams and the 
classical limit of the eikonal amplitude gives the required post-Minkowskian equation. The novelty of the recursive N\"other coupling is that we do not 
require to assume any quantum theory and quantum amplitude which does not have any closed form except in the eikonal limit. 
The post-Minkowskian field equations come just from identifying the self-interaction (self-interaction only means the coupling of the N\"other current with 
the field variable for a specific classical Lagrangian) as a source of next order field equation at the classical level.

\section{Conclusion}
We have derived the post-Minkowskian series of GR up to the third order term by recursively adding the self-interactions starting from the projected action of linearized gravity. The whole thing is done without any reference to the quantum theory which needs eikonal approximation to derive the post-Minkowskian series.

The recursive N\"other coupling for Pauli-Fierz action of spin 2 field in flat spacetime does not stop after any finite terms. The end result is an infinite series which gives GR with some limitations. The diffeomorphism invariance of GR (in other words gauge invariance of GR) makes some components of metric redundant. In our way of doing recursive N\"other coupling in the projected Fierz-Pauli action has the potential to generate a redundancy free theory of GR. This may be interesting in solving Einstein equations in a very complicated system, such as in numerical relativity. However unlike usual GR, the diffeomorphism invariance is not obvious in such theory. Clearly, more research is necessary in this direction.

In the first-order formulation, general relativity in the linearized approximation is done in only one recursion step \cite{Deser}. Hence the corresponding self-interaction consists of the complete series of the post-Minkowskian expansion. Another interesting thing to see is, whether the correspondence just shown between the post-Minkowskian series and the recursive addition of the self-interaction, works for higher curvature theories of gravity. In the view of \cite{Mod}, the answer is no, at least for quadratic gravity. This is because, the linearized version of the quadratic gravity does not contain any second derivative of the linearized metric field whereas its post-Minkowskian expansion counterpart certainly has the second derivative of the expanded metric.

As already mentioned, this approach of recursive N\"other coupling of the energy-momentum tensor to the projected massless spin 2 field in a flat spacetime background, despite reservations from some quarters, clearly acquires the status of applicability to gravitational wave signal processing in an autonomous manner, without  having to consider weak-field approximations to general relativity. As more and more observations are made by extant and planned laser-interferometer observatories, one can indeed avoid complications associated with general relativity, and directly proceed to use this approach for analyses of observational data. Clearly, the method has obvious shortcomings when analyses of the spacetime in close proximity of coalescence of compact binary is required, but that is indeed another story.

With regard to the central question not addressed in this work, namely how best to do the recursive N\"other coupling to derive full general relativity, one must mention a particular complication of formulations of general relativity as a gauge theory. Clearly, the {\it gauge} group of general relativity is not the local Poincar\'e group, that is not large enough to include the group of spacetime diffeomorphisms. Rather, this is perhaps best described as the {\it semi-direct} product of the group of diffeomorphisms and the local Lorentz group. In the standard metric formalism of general relativity, local Lorentz invariance is not manifest, because only diffeomorphism invariance is used to construct the theory. Yet, for coupling spinors to spacetime, local Lorentz invariance is crucial. The challenge is to bring in manifest local Lorentz invariance through the method of recursive N\"other coupling. 

\section{Acknowledgement}

SKR acknowledges Pratyusava Baral of Presidency University, Kolkata for the helpful discussion and DST-Inspire grant for financial support during the period that this work was done.

\end{document}